\documentclass[aps,twocolumn]{revtex4}

\def\vb#1{\mbox{\boldmath$#1$}}
\def\pd#1#2{\frac{\partial #1}{\partial #2}}

\def\bdot{\,\vb{\cdot}\,}

\def\cal#1{\mathcal{#1}}

\newcommand{\bc}{\begin{center}}
\newcommand{\ec}{\end{center}}
\newcommand{\bt}{\begin{tabbing}}
\newcommand{\et}{\end{tabbing}} 
\newcommand{\be}{\begin{eqnarray*}}
\newcommand{\ee}{\end{eqnarray*}}
\newcommand{\bs}{\begin{slide}}
\newcommand{\es}{\end{slide}}

\begin{document}

\title{On the proper choice of a Lorentz-covariant relativistic Lagrangian}

\author{Alain J.~Brizard}
\affiliation{Department of Chemistry and Physics, Saint Michael's College, Colchester, VT 05439, USA} 

\begin{abstract}
A proper choice for the sign of the Lorentz-covariant relativistic Lagrangian for a particle moving in an electromagnetic field is presented. This choice is based on which space-time metric is used in the representation of the relativistic particle dynamics.
\end{abstract}

\begin{flushright}
December 1, 2009
\end{flushright}

%\pacs{52.30.Gz, 52.65.Tt}

\maketitle

The relativistic Lagrangian for a particle moving in an electromagnetic field is defined up to an overall sign. In almost all textbooks, the Lagrangian is defined with an overall negative sign \cite{L_L,Jackson,Goldstein,Corben_Stehle}, while in some, the Lagrangian is defined with a positive sign 
\cite{Lanczos,Gray_etal}. In the present brief note, we show that the overall sign of the Lorentz-covariant relativistic Lagrangian is determined consistently by the choice of space-time metric used in describing the particle dynamics:
\begin{equation}
{\sf g} \;\equiv\; \left( \begin{array}{cccc}
\sigma & 0 & 0 & 0 \\
0 & -\,\sigma & 0 & 0 \\
0 & 0 & -\,\sigma & 0 \\
0 & 0 & 0 &-\,\sigma  \\
\end{array} \right),
\label{eq:g_def}
\end{equation}
where $\sigma = +\,1$ for a time-like metric and $\sigma = -\,1$ for a space-like metric. 

First, we consider the following relativistic Lagrangian describing the motion of a particle (of mass $m$ and charge $e$) under the influence of an electromagnetic field
\begin{equation}
L_{\epsilon} \;\equiv\; \epsilon \left[\; mc^{2}\,\gamma^{-1} \;+\; e \left( \Phi \;-\; {\bf A}\bdot\frac{\dot{{\bf x}}}{c} \right) \;\right],
\label{eq:Lag_epsilon}
\end{equation}
where $\epsilon = \pm\,1$ and $A^{\mu} = (\Phi, {\bf A})$ denotes the components of the electromagnetic four-potential. Here, the relativistic factor is \begin{equation}
\gamma \;=\; 1/\sqrt{1 - |\dot{{\bf x}}|^{2}/c^{2}},
\label{eq:gamma}
\end{equation} 
and a dot refers to a time derivative. The choice $\epsilon = -1$ is most commonly adopted because the regular Hamiltonian $H$ is defined by regular Legendre transformation as
\begin{equation}
H \;\equiv\; \dot{{\bf x}}\bdot\pd{L_{-}}{\dot{{\bf x}}} \;-\; L_{-} \;=\; m\gamma\,c^{2} \;+\; e\;\Phi,
\label{eq:H_rel}
\end{equation}
while the classical Lagrangian
\begin{equation}
L_{\rm cl} \;=\; \left( m\,{\bf v} \;+\; \frac{e}{c}\,{\bf A}\right)\bdot\dot{{\bf x}} \;-\; \left( \frac{m}{2}\,|{\bf v}|^{2} \;+\; e\,\Phi\right)
\label{eq:nonRel_Lag}
\end{equation}
is obtained from the relativistic Lagrangian $L_{-}$ in the nonrelativistic limit $(|\dot{{\bf x}}| \ll c)$, where the constant $-\,mc^{2}$ was omitted from Eq.~(\ref{eq:nonRel_Lag}). Here, the identity ${\bf v} \equiv \dot{{\bf x}}$ for the particle velocity results from the Euler-Lagrange equation for ${\bf v}$ (since $\partial L_{\rm cl}/\partial\dot{{\bf v}} \equiv 0$). 

Next, we show that the choice for $\epsilon$ for a Lorentz-covariant relativistic Lagrangian can be consistently based on the sign $\sigma$ of the space-time metric (\ref{eq:g_def}). For this purpose, we introduce the notation
\begin{equation}
A^{0}\,B^{0} \;-\; {\bf A}\bdot{\bf B} \;\equiv\; \sigma\;A^{\mu}\,B_{\mu} \;=\; \sigma\;A^{\mu}\,g_{\mu\nu}\,B^{\nu},
\label{eq:metric_def}
\end{equation}
where $A^{\mu} = (A^{0}, {\bf A})$ and $B^{\mu} = (B^{0}, {\bf B})$ are two arbitrary four-vectors and we used the space-time metric (\ref{eq:g_def}). Hence, the relativistic factor (\ref{eq:gamma}) is defined covariantly as
\begin{equation}
\gamma \;\equiv\; c\,(\sigma\,\dot{x}_{\mu}\,\dot{x}^{\mu})^{-1/2}.
\label{eq:gamma_4}
\end{equation} 
The relativistic Lagrangian (\ref{eq:Lag_epsilon}) can be written in Lorentz-covariant form as
\begin{eqnarray}
{\cal L}_{\epsilon} & \equiv & \epsilon\,mc\;\sqrt{(\dot{x}^{0})^{2} \;-\; |\dot{{\bf x}}|^{2}} \;+\; \epsilon\;\frac{e}{c} \left( \Phi\,\dot{x}^{0} \;-\;
{\bf A}\bdot\dot{{\bf x}} \right) \nonumber \\
 & = & \epsilon \left( mc\;\sqrt{\sigma\,\dot{x}_{\mu}\,\dot{x}^{\mu}} \;+\; \frac{e}{c}\;\sigma\,A_{\mu}\dot{x}^{\mu} \right) 
\label{eq:Lag_epsilon_4}
\end{eqnarray}
where $\dot{x}^{0} = c$. The action integral 
\begin{equation}
S_{\epsilon} \;=\; \int\;{\cal L}_{\epsilon}\,dt \;\equiv\; \epsilon\;\int \left( mc\,ds \;+\; \frac{e}{c}\;\sigma\,A_{\mu}\,dx^{\mu} \right)
\label{eq:S_def}
\end{equation}
therefore involves an integration involving two Lorentz invariants $mc\,ds$ and $(e/c)\,A_{\mu}\,dx^{\mu}$, where $ds^{2} \equiv \sigma\,dx_{\mu}
dx^{\mu}$ and used the definition (\ref{eq:metric_def}). The Euler-Lagrange equations 
\[ \frac{d}{dt} \left( \pd{{\cal L}_{\epsilon}}{\dot{x}^{\mu}} \right) \;=\; \pd{{\cal L}_{\epsilon}}{x^{\mu}} \]
obtained from the relativistic Lagrangian (\ref{eq:Lag_epsilon_4}) or the variational principle based on the action integral (\ref{eq:S_def}), are expressed as
\begin{equation} 
\frac{d}{dt} \left( mc\,\gamma\,\frac{dx_{\mu}}{dt} \;+\; \frac{e}{c}\;A_{\mu} \right) \;=\; \frac{e}{c} \left( \pd{A_{\nu}}{x^{\mu}}\;
\frac{dx^{\nu}}{dt} \right), \label{eq:EL_mu} 
\end{equation}
where the common factor of $\epsilon\,\sigma$ has been removed from both sides [i.e., the Euler-Lagrange equations (\ref{eq:EL_mu}) are independent of the choice of the sign $\epsilon$ of the Lagrangian and the sign $\sigma$ of the space-time metric]. If we now define the proper time derivative 
$d/d\tau \equiv \gamma\,d/dt$ and the four-velocity $u^{\mu} \equiv dx^{\mu}/d\tau$, the Euler-Lagrange equations (\ref{eq:EL_mu}) become (upon multiplication by $\gamma$)
\begin{equation}
m\;\frac{du_{\mu}}{d\tau} \;=\; \frac{e}{c} \left( \pd{A_{\nu}}{x^{\mu}} \;-\; \pd{A_{\mu}}{x^{\nu}} \right) u^{\nu} \;\equiv\; \frac{e}{c}\,
F_{\mu\nu}\;u^{\nu}.
\label{eq:u_tau}
\end{equation}
Note that these equations can be obtained directly from the variational principle $\delta\int\pounds_{\epsilon}\,d\tau = 0$, where
\begin{equation}
\pounds_{\epsilon} \;\equiv\; \gamma\,{\cal L}_{\epsilon} \;=\; \epsilon \left( mc\,\sqrt{\sigma\,u_{\mu}u^{\mu}} \;+\; \frac{e}{c}\,\sigma\,A_{\mu}\,
u^{\mu} \right).
\label{eq:calL_epsilon}
\end{equation}
The associated Euler-Lagrange equations
\[ \frac{d}{d\tau} \left( \pd{\pounds_{\epsilon}}{u^{\mu}} \right) \;=\; \pd{\pounds_{\epsilon}}{x^{\mu}} \]
directly yield the covariant equations of motion (\ref{eq:u_tau}).

The covariant and contravariant components of the canonical four-momentum are derived from the Lagrangian (\ref{eq:Lag_epsilon_4}) as
\begin{eqnarray}
p_{0} & \equiv & \pd{{\cal L}_{\epsilon}}{\dot{x}^{0}} \;=\; \epsilon \left( m\,\gamma\,\dot{x}^{0} \;+\; \frac{e}{c}\,\Phi \right) \;\equiv\; \sigma\,
p^{0}, 
\label{eq:p_0_def} \\
p_{i} & \equiv & \pd{{\cal L}_{\epsilon}}{\dot{x}^{i}} \;=\; -\,\epsilon \left( m\,\gamma\,\dot{x}^{i} \;+\; \frac{e}{c}\,A^{i} \right) \;\equiv\; 
-\,\sigma\,p^{i},
\label{eq:p_i_def}
\end{eqnarray}
where we used the space-time metric (\ref{eq:g_def}) to relate the covariant and contravariant components $p_{\mu} \equiv g_{\mu\nu}\,p^{\nu}$. A consistent choice for the sign $\epsilon$ of the relativistic Lagrangian (\ref{eq:Lag_epsilon_4}) [and Eq.~(\ref{eq:calL_epsilon})] is $\epsilon \equiv \sigma$ and, therefore, the proper choice for the Lorentz-covariant relativistic Lagrangian for a particle moving in an electromagnetic field is
\begin{equation}
{\cal L}_{\sigma} \;\equiv\; \sigma\,mc\;\sqrt{\sigma\,\dot{x}_{\mu}\dot{x}^{\mu}} \;+\; \frac{e}{c}\;A_{\mu}\,\dot{x}^{\mu},
\label{eq:Lag_sigma}
\end{equation}
or
\begin{equation}
\pounds_{\sigma} \;\equiv\; \sigma\,mc\;\sqrt{\sigma\,u_{\mu}u^{\mu}} \;+\; \frac{e}{c}\;A_{\mu}\,u^{\mu}.
\label{eq:pounds_sigma}
\end{equation}
Hence, we see that the choice of sign for the relativistic Lagrangian (\ref{eq:Lag_sigma}) is intimately connected to the choice of space-time metric. We note that the components of the canonical four-momentum
\begin{equation}
p^{\mu} \;\equiv\; \left\{ \begin{array}{rcl}
\partial{\cal L}_{\sigma}/\partial\dot{x}_{\mu} & = & m\,\gamma\;\dot{x}^{\mu} \;+\; (e/c)\;A^{\mu} \\
 &  & \\
\partial\pounds_{\sigma}/\partial u_{\mu} & = & m\,u^{\mu} \;+\; (e/c)\;A^{\mu}
\end{array} \right.
\label{eq:p_mu}
\end{equation}
are naturally independent of the choice ($\sigma = \pm\,1$) of the space-time metric (\ref{eq:g_def}).

Lastly, the Lagrangian (\ref{eq:Lag_sigma}) satisfies the identity
\begin{equation} 
\dot{x}^{\mu}\;\pd{{\cal L}_{\sigma}}{\dot{x}^{\mu}} \;=\; \left[\; m\,\gamma\;\left(\dot{x}^{\mu}\,
\dot{x}_{\mu}\right) \;+\; \frac{e}{c}\;A_{\mu}\,\dot{x}^{\mu} \;\right] \;\equiv\; {\cal L}_{\sigma}, 
\label{eq:calH_sigma}
\end{equation}
where we used the definition (\ref{eq:gamma}) for the relativistic factor $\gamma$. Using the identity (\ref{eq:calH_sigma}), the extended Hamiltonian 
${\cal H}_{\sigma}$ derived from the extended Legendre transformation 
\begin{equation}
{\cal H}_{\sigma} \;\equiv\; p_{\mu}\,\dot{x}^{\mu} \;-\; {\cal L}_{\sigma} 
\label{eq:cal_H}
\end{equation}
therefore vanishes identically. 

In summary, while the choice for the sign $\epsilon$ of the relativistic Lagrangian (\ref{eq:Lag_epsilon}) does not impact its associated Euler-Lagrange equations, the choice $\epsilon = -1$ is commonly used because of its natural connection with the relativistic Hamiltonian (\ref{eq:H_rel}) and its non-relativistic limit (\ref{eq:nonRel_Lag}). When a Lorentz-covariant relativistic Lagrangian formulation is adopted, however, a consistent choice for 
$\epsilon \equiv \sigma$ is associated with the sign $\sigma$ of the space-time metric (\ref{eq:g_def}) used.

\end{document}